\documentclass[english,prl,superscriptaddress,reprint]{revtex4-1}
\usepackage[LGR,T1]{fontenc}
\usepackage[latin9]{inputenc}
\usepackage{textcomp}
\usepackage{amstext}
\usepackage{graphicx}
\usepackage{subscript}

\makeatletter

\DeclareRobustCommand{\greektext}{%
  \fontencoding{LGR}\selectfont\def\encodingdefault{LGR}}
\DeclareRobustCommand{\textgreek}[1]{\leavevmode{\greektext #1}}
\ProvideTextCommand{\~}{LGR}[1]{\char126#1}

\usepackage{amssymb,amsmath}
\usepackage{hyperref}

\makeatother

\usepackage{babel}
\begin{document}

\title{Microanalysis of single-layer hexagonal boron nitride islands on
Ir(111)}

\author{Marin Petrovi\'{c}}
\email{marin.petrovic@gast.uni-due.de}

\altaffiliation{Permanent address: Center of Excellence for Advanced Materials and Sensing Devices, Institute of Physics, Bijenika 46, 10000 Zagreb, Croatia}

\affiliation{Faculty of Physics and CENIDE, University of Duisburg-Essen, Lotharstr. 1,
D-47057 Duisburg, Germany}

\author{Ulrich Hagemann}

\address{Interdisciplinary Center for Analytics on the Nanoscale (ICAN), Nano
Energie Technik Zentrum (NETZ), Carl-Benz-Str. 199, D-47047 Duisburg,
Germany}

\author{Michael Horn-von Hoegen}

\address{Faculty of Physics and CENIDE, University of Duisburg-Essen, Lotharstr. 1,
D-47057 Duisburg, Germany}

\author{Frank-J. Meyer zu Heringdorf}

\address{Faculty of Physics and CENIDE, University of Duisburg-Essen, Lotharstr. 1,
D-47057 Duisburg, Germany}
\begin{abstract}
Large hexagonal boron nitride (hBN) single-layer islands of high crystalline
quality were grown on Ir(111) via chemical vapor deposition (CVD)
and have been studied with low-energy electron microscopy (LEEM).
Two types of hBN islands have been observed that structurally differ
in their shape and orientation with respect to iridium, where the
former greatly depends on the iridium step morphology. Photoemission
electron microscopy (PEEM) and IV-LEEM spectroscopy revealed that
the two island types also exhibit different work functions and bindings
to iridium, which provides an explanation for differences in their
shape and growth modes. In addition, various temperatures were used
for the CVD synthesis of hBN, and it was found that at temperatures
higher than $\approx950$ \textdegree C boron atoms, originating either
from decomposed borazine molecules or disintegrated hBN islands, can
form additional compact reconstructed regions. The presented results
are important for advancement in synthesis of high-quality hBN and
other boron-based layered materials, and could therefore expedite
their technological implementation.
\end{abstract}

\keywords{hexagonal boron nitride, iridium, LEEM, PEEM}
\maketitle

\section{Introduction}

Alongside pioneering graphene (G), researchers have characterized
a large number of other two-dimensional (2D) materials \cite{Xu2013}
that may potentially be used in novel applications. Among them, single-layer
hexagonal boron nitride (hBN) is particularly interesting due to the
isostructurality to G \cite{Pakdel2014} and its insulating nature,
i.e., a large bandgap of $\sim6$ eV \cite{Cassabois2016}. These
properties, alongside with atomic flatness, chemical inertness and
the absence of charge traps, designate hBN as a very important 2D
material, ideal for the formation of heterostructures with G that
promise various technological advancements, for example, in transistor
\cite{Dean2010,Bresnehan2012} or integrated circuits \cite{Levendorf2012}
fabrication. Besides in combination with graphene, hBN may be used
for encapsulation of other 2D materials which enhances their performance
in various devices \cite{Lee2015enc}. Furthermore, other applications
of hBN include use as an active laser medium \cite{Watanabe2004}
and an ultra-thin insulating support for molecular self-assembly \cite{Corso2004b,Schulz2013}.

Single layers of hBN can be produced via chemical vapor deposition
(CVD) on various single-crystal metal substrates \cite{Auwarter1999a,Morscher2006,Goriachko2007,Cavar2008,Dong2010,Muller2010,Orlando2011,Joshi2012}.
Similar to the case of epitaxial G \cite{Wintterlin2009a,Celasco2017},
some metals are characterized by strong interaction with hBN, such
as Ru(0001) or Rh(111), while on other substrates hBN is more weakly
bound, e.g., on Pt(111) or Cu(111) \cite{Preobrajenski2005,Preobrajenski2007a,Laskowski2010}.
Optimal hBN production requires a substrate to which, on the one hand,
hBN binds sufficiently strongly to enable the growth of uniformly
oriented domains and on the other hand, does not bind too strongly
in order to preserve its intrinsic properties. These requirements
are fulfilled for hBN on Ir(111) {[}hBN/Ir(111){]} because there the
interaction is characterized as intermediate \cite{Preobrajenski2007,Laskowski2008}.
Moreover, high-quality G samples with preserved intrinsic electronic
structure are easily synthesized on Ir(111) \cite{Kralj2011} and
are a very good starting point for the fabrication of G-based heterostructures.
Therefore, Ir(111) is a suitable substrate for studying both fundamental
properties of hBN as well as more complex systems comprised of hBN
and other 2D materials.

A number of studies of hBN/Ir(111) have been conducted so far, but
there are still aspects of the system that require further investigation.
For example, hBN exhibits two preferential orientations with respect
to Ir(111) which were identified via X-ray photoelectron diffraction
(XPD) \cite{Orlando2014} and scanning tunneling microscopy (STM)
\cite{FarwickzumHagen2016}, but a more detailed morphological characterization
of the respective hBN domains is lacking, especially on the micro-scale.
Also, differences in the electronic structure of the two hBN orientations
are not known, since only area-averaging ARPES has been performed
for valence band mapping \cite{Usachov2012}. In addition, the above
mentioned studies report on the decomposition of hBN at elevated temperatures
and emergence of other boron species on the iridium surface, as evidenced
by STM imaging \cite{FarwickzumHagen2016} as well as disappearance
and appearance of different B 1s peaks in XPS \cite{Usachov2012}.
Details of these processes, which are vital for establishment of high-quality
hBN synthesis, are not known. Moreover, insight into formation of
various boron structures could initiate the production of borophene\textendash an
exciting 2D boron analogue of G \cite{Mannix2015}\textendash on iridium.
To investigate all these issues, the use of a complementary microscopic
technique is necessary. Low-energy electron microscopy (LEEM) is an
ideal tool for this task because it enables \textit{in situ}, real-time
observation of growth-related phenomena with the spatial resolution
in the nanometer range. With this method, hBN has been characterized
on e.g., Pt(111) \cite{Cavar2008,Zhang2015c}, Ru(0001) \cite{Goriachko2008,Sutter2011a}
and Co(0001) \cite{Orofeo2013}, but similar studies addressing hBN/Ir(111)
are still lacking. The aim of this work is to provide first microscopic
insights into this epitaxial system primarily via LEEM and in conjunction
with several other methods, which will supplement our findings. In
this way we tackle challenges of CVD synthesis of hBN, which is the
most promising technique for the large-scale production of hBN.

\section{Experimental}

Growth of hBN was conducted in an ultra-high vacuum (UHV) setup with
a base pressure better than $3\times10^{-10}$ mbar. CVD was used
for the synthesis of hBN during which a hot (111) surface of Ir single-crystal
(Mateck) was exposed to borazine (B\textsubscript{3}H\textsubscript{6}N\textsubscript{3},
Katchem). The crystal was cleaned by Ar\textsuperscript{+} sputtering
at 2 keV followed by several cycles of heating in oxygen at 900 \textdegree C
and annealing at 1200 \textdegree C. In the case when only hBN had
to be removed from the surface, heating in oxygen and annealing were
sufficient. The sample temperature was measured with an infrared pyrometer.
Borazine was kept in a home-built Peltier cooler, which enabled realization
of the sub-zero temperatures needed to prevent borazine decomposition.
Thus, borazine was available at all times at the UHV setup, and prior
to each hBN synthesis the borazine inlet line was pumped down to remove
any decomposition residues. Unless otherwise noted, borazine pressure
during CVD was $10^{-8}$ mbar.

LEEM and PEEM measurements were carried out \textit{in situ} with
an Elmitec SPE-LEEM III microscope at the University of Duisburg-Essen.
PEEM measurements were performed with the aid of an Hg discharge lamp
providing $\sim4.9$ eV photons that were relevant for our samples.
The spatial resolution of the microscope is $\sim10$ nm. Atomic force
microscope (AFM) measurements were performed \textit{ex situ} in air
with a Veeco Dimension 3100 microscope operated in tapping mode. An
Auger Nanoprobe 710 from Ulvac-Phi located at the ICAN center in Duisburg
was used for \textit{ex situ} Auger electron spectroscopy (AES) and
scanning electron microscopy (SEM) with a base pressure of $6\times10^{-10}$
mbar. Primary electron energy for AES was 10 keV and 5 keV for SEM.

\section{Results and discussion}

\subsection{Temperature-dependent growth}

\begin{figure}
\begin{centering}
\includegraphics{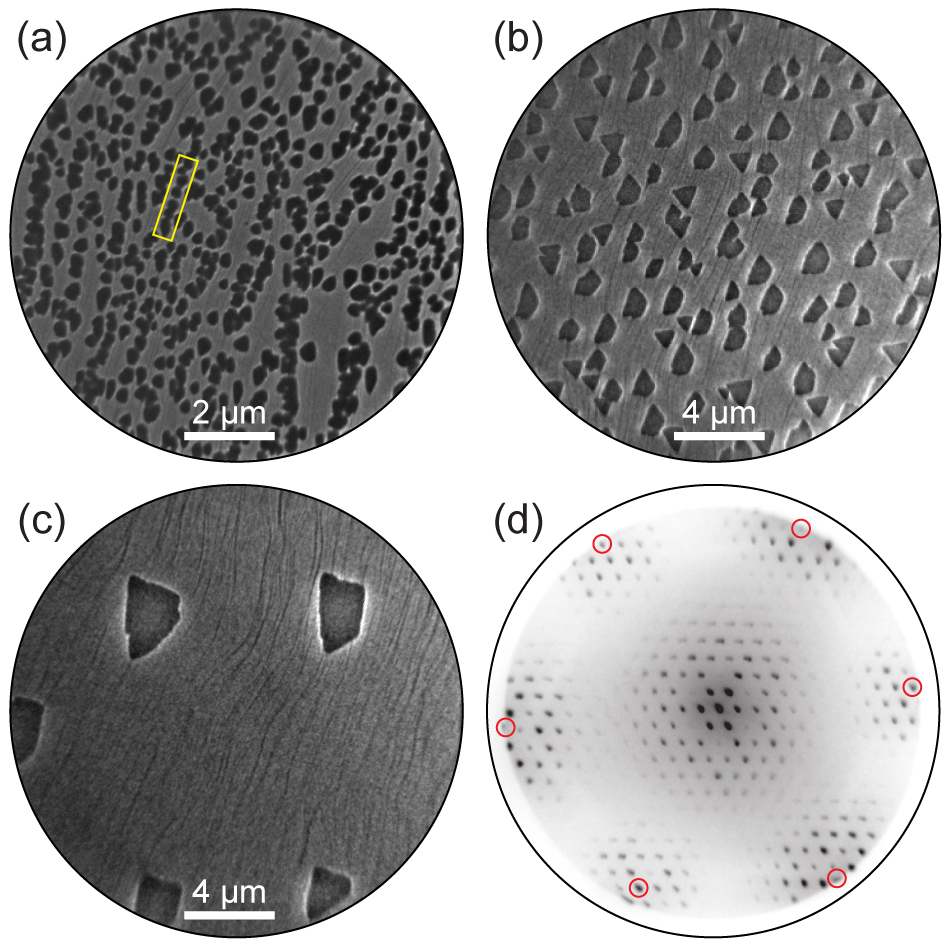}
\par\end{centering}
\caption{\label{fig1}(a)-(c) LEEM images of hBN islands synthesized at 800,
900 and 1100 \textdegree C after 100 s of borazine dosing at $5\times10^{-8}$,
$5\times10^{-8}$ and $3\times10^{-8}$ mbar, respectively. Images
are recorded at the given synthesis temperature. E = 17 eV. (d) Characteristic
LEED pattern originating from several coalescing hBN islands synthesized
at 900 \textdegree C. Red circles indicate positions of first-order
hBN spots. The image was recorded at room temperature. E = 31 eV.}
\end{figure}

In order to investigate different growth conditions, hBN islands were
synthesized at various temperatures in LEEM. Fig. \ref{fig1}(a)-(c)
show characteristic island morphologies for the growth temperatures
of 800, 900 and 1100 \textdegree C (note that fields of view are not
the same in all panels). In all cases, island nucleation was initiated
at Ir step edges. At low synthesis temperatures, this resulted in
the formation of strings of hBN islands that decorate the same Ir
step edge, as marked in Fig. \ref{fig1}(a) with a yellow rectangle.
The corresponding island densities in Fig. \ref{fig1}(a)-(c) are
8.09, 0.58 and 0.02 \textgreek{m}m\textsuperscript{-2}, thus spanning
several orders of magnitude. At the same time, the average island
size increases with synthesis temperature. These findings are in line
with published STM data of hBN synthesized at different temperatures
\cite{FarwickzumHagen2016} and the temperature- and dose-dependence
of the classical nucleation theory in general. With the growth conditions
we used, hBN islands with sizes up to $\sim10$ \textgreek{m}m (for
1100 \textdegree C synthesis) were readily obtained. During prolonged
exposure to borazine, individual islands coalesce, resulting in a
full coverage of hBN and suppression of multilayer growth, similar
to hBN growth on Ru(0001) \cite{Sutter2011a}.

Upon cooldown to room temperature, the formation of wrinkles was not
observed on either islands or a full hBN monolayer {[}see Fig. \ref{fig4}(a)
or (d) as an example{]}. However, wrinkling was found before for hBN
on weakly interacting substrates, such as Pt(111) \cite{Zhang2015c}
or Cu foils \cite{Kim2011a}. In contrast to hBN, G wrinkles on Ir(111)
are routinely observed via LEEM and have been characterized in detail
\cite{NDiaye2009,Petrovic2015}. Therefore, our data indicates that
wrinkling does not occur on hBN/Ir(111). This suggests that (i) the
binding of hBN and G to Ir is different and/or that (ii) the elastic
stiffness of hBN is lower than that of G. In comparison, theoretical
calculations give binding strength to Ir of 138 meV per G and 170.8
meV per hBN unit cell \cite{FarwickzumHagen2016}. Also, AFM indentation
experiments of suspended layers provided an elastic stiffness value
of 340 Nm\textsuperscript{-1} for G \cite{Lee2008} and, in combination
with theoretical modeling, 292 Nm\textsuperscript{-1} for hBN \cite{Song2010}.
Because binding and stiffness values are comparably close, it seems
that a combination of both contributes to the absence of wrinkle formation
in hBN.

A characteristic low-energy electron diffraction (LEED) pattern of
several coalescing hBN islands synthesized at 900 \textdegree C is
shown in Fig. \ref{fig1}(d). The most prominent features of the pattern
are diffraction spots that reflect the presence of the moiré structure
of hBN/Ir(111) arising from the mismatch of the two surface lattices.
The high quality of our hBN islands is confirmed by a large number
of sharp moiré spots, which range up to the sixth order and thus span
almost the entire first Brillouin zone. A closer look at the diffraction
spots in Fig. \ref{fig1}(d) reveals that the hBN area under examination
consists of several slightly rotated domains (see Subsection 3.2.
for details on rotational registry of hBN and Ir). From a more detailed
analysis of similar LEED patterns, we calculate the ratio of hBN and
Ir reciprocal lattice vector modules, $k_{\textrm{hBN}}/k_{\textrm{Ir}}$,
to be $\left(1.09\pm0.01\right)$. Hence, the structure of hBN on
Ir is such that $\sim10.9$ unit cells of hBN are superposed on top
of 10 unit cells of Ir. The calculated ratio of 1.09 matches the ratio
of $11.7/10.7$ (and also $12/11$ as the best commensurate approximation),
which was recently reported as an exact description of the hBN/Ir(111)
moiré structure \cite{FarwickzumHagen2016}. Knowing that the Ir real
space lattice constant is $a_{\textrm{Ir}}=2.715$ Å \cite{Arblaster2010},
we calculate the lattice constant of hBN to be $a_{\textrm{hBN}}=\left(2.50\pm0.02\right)$
Å, which is in good agreement with the published values for hBN/Ir(111)
of 2.483 Å \cite{FarwickzumHagen2016} as well as that for bulk hBN
of 2.505 Å \cite{Paszkowicz2002}. Also, the $k_{\textrm{hBN}}/k_{\textrm{Ir}}$
ratio enables calculation of the corresponding moiré periodicity,
which turns out to be $a_{\textrm{moiré}}=\left(30\pm3\right)$ Å
(for the hBN aligned to Ir), being consistent with the reported STM
measurements \cite{Schulz2014,FarwickzumHagen2016}.

\begin{figure}
\begin{centering}
\includegraphics{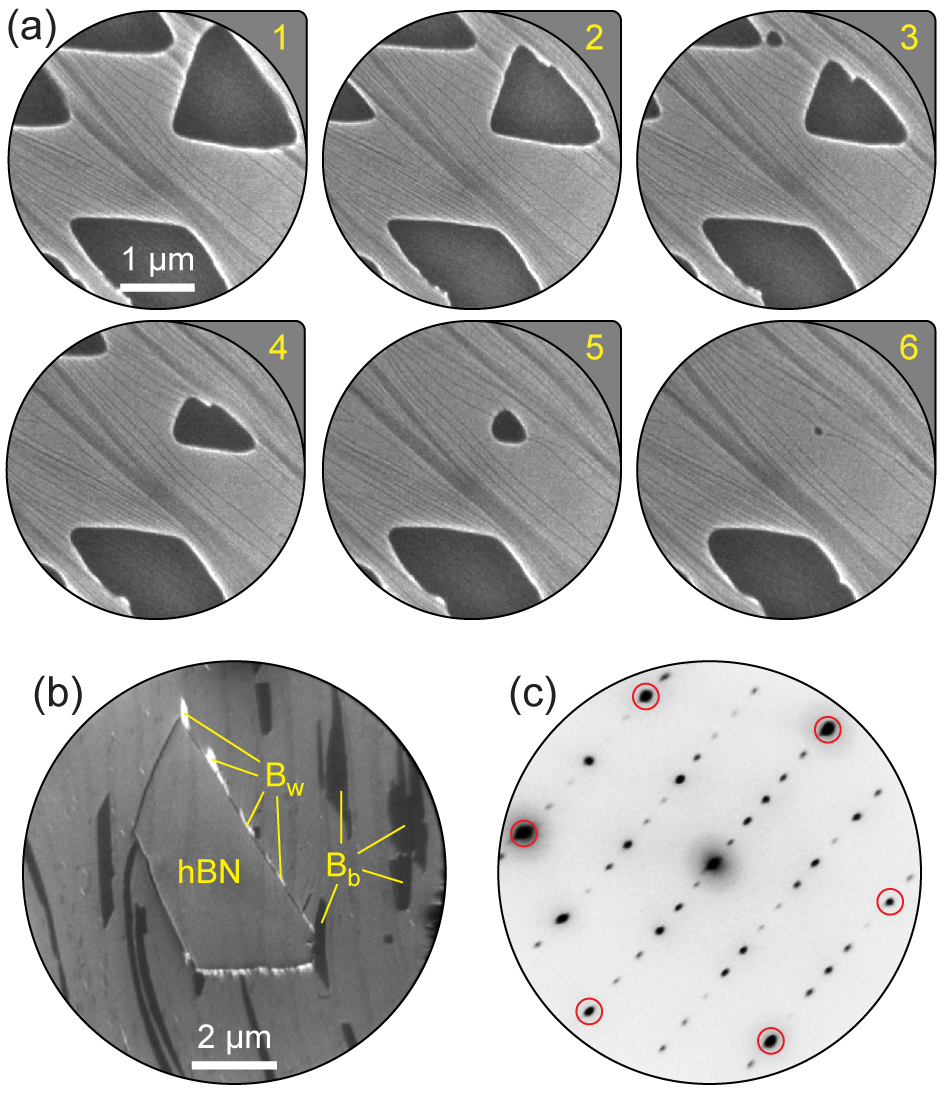}
\par\end{centering}
\caption{\label{fig2}(a) Decomposition of hBN islands at 1100 \textdegree C.
LEEM images are taken in equidistant time intervals of 53 s. E = 16.6
eV. (b) hBN island surrounded by additional boron in the form of dark
regions filling Ir terraces (B\protect\textsubscript{b}) and white
patches decorating hBN island edges (B\protect\textsubscript{w}).
Synthesis temperature was 1100 \textdegree C, image recorded at room
temperature. E = 16.5 eV. (c) \textgreek{m}-LEED pattern of a B\protect\textsubscript{b}
region exhibiting a $\left(6\times2\right)$ superstructure with respect
to Ir. Red circles indicate positions of first-order Ir spots. E =
43 eV.}
\end{figure}

The stability of hBN islands was examined by keeping the sample at
high temperatures after stopping the borazine dosage. It was found
that in the case of sufficiently high temperatures, synthesized hBN
islands disintegrate from the perimeter inward and eventually vanish
from the field of view. An example of disintegration is shown in a
sequence of LEEM images in Fig. \ref{fig2}(a) for the sample temperature
of 1100 \textdegree C. There are two possible routes for the B and
N atoms after an hBN island decomposes: the atoms can desorb from
the sample or they can remain on the Ir surface, either in the adsorbed
or dissolved form. N desorption and B dissolution into the bulk seem
to be the most probable scenario because N\textsubscript{2} molecules
desorb from the (111) surface of Ir even below room temperature \cite{Cornish1990},
while iridium borides of various stoichiometric ratios can be formed
at high temperatures \cite{Zeiringer2015}. Further experiments corroborating
this assumption are described in the following.

Fig. \ref{fig2}(b) shows the surface of the sample after hBN synthesis
at 1100 \textdegree C and subsequent cooldown to room temperature.
Beside a compact hBN island, extra material is visible in the form
of dark, elongated features that predominantly follow the shape of
Ir step edges, and irregular bright patches that decorate hBN island
edges. We label this additional material as B\textsubscript{b} and
B\textsubscript{w}, respectively. B\textsubscript{b} regions appeared
during cooldown at $\approx950$ \textdegree C, and they were also
observed if cooldown was performed before any hBN nucleation or after
hBN disintegration. We noticed that B\textsubscript{w} regions grew
if the sample was kept at elevated synthesis temperature without dosing
borazine and the hBN island they grew attached to remained unaltered.
All this evidence suggests that the B\textsubscript{b} and B\textsubscript{w}
regions are in fact reconstructed boron areas. B atoms are fed to
those regions from the borazine gas phase and from disintegrated hBN
islands, while the mechanism of B\textsubscript{b} and B\textsubscript{w}
formation is either (i) diffusion of B atoms across the Ir surface
and eventual nucleation or (ii) dissolution of B atoms into Ir subsurface
regions followed by segregation due to the decrease of solubility
at lower temperatures. Therefore, the threshold temperature for significant
disintegration of hBN on (111) face of Ir and the diffusion of B into
the subsurface regions is $\approx950$ \textdegree C, and a slightly
lower temperature (900 \textdegree C) was used for synthesis when
hBN islands without excessive B regions surrounding them were desired.
\textgreek{m}-LEED analysis, enabling diffraction studies of sub-micrometer-sized
areas on the sample surface, revealed that B\textsubscript{b} regions
exhibit an adsorbate (not adlayer) form due to the relatively dilute
$\left(6\times2\right)$ superstructure {[}see the diffraction pattern
in Fig. \ref{fig2}(c){]}, and they were present on the surface in
three equivalent 120\textdegree -rotated variants. B\textsubscript{w}
regions exhibit a $\left(1\times1\right)$ (or possibly disordered)
structure (not shown). We also note that B\textsubscript{b} and B\textsubscript{w}
regions are easily etched away at elevated temperatures with oxygen.

\begin{figure}
\begin{centering}
\includegraphics{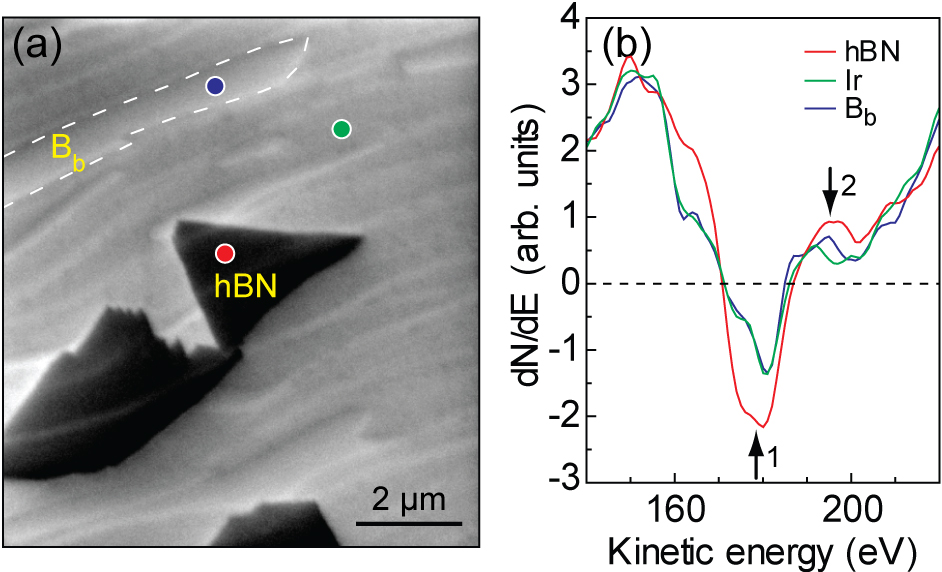}
\par\end{centering}
\caption{\label{fig3}(a) SEM image of hBN islands (dark) on Ir(111) with elongated
B\protect\textsubscript{b} regions surrounding them (bright, one
of the islands is indicated by a dashed line). Colored dots mark the
positions where Auger spectra shown in (b) were recorded. (b) Auger
spectra of an hBN island, bare Ir and B\protect\textsubscript{b}
region. Arrows indicate regions that enable identification of B on
the sample (see text for details).}
\end{figure}

In order to confirm our speculations on the chemical composition of
B\textsubscript{b} and B\textsubscript{w} regions, AES was performed.
A suitable area for AES was first found with SEM and is shown in Fig.
\ref{fig3}(a). hBN islands appear dark and B\textsubscript{b} regions
are visible as elongated, slightly lighter structures when compared
with the surrounding bare Ir. The mode of contrast generation in SEM
(which is different to the one in LEEM) does not enable imaging of
the small B\textsubscript{w} regions attached to hBN. Because of
this, B\textsubscript{w} regions were not accessible for Auger analysis.
Three characteristic positions (bare Ir, hBN and B\textsubscript{b})
marked by colored dots (green, red and blue, respectively) were chosen
and the resulting AES spectra are shown in Fig. \ref{fig3}(b), with
a focus on the energy range where the main B peak is expected to be
found \cite{Sutter2014a}. A notable Ir peak is located in the same
region and contributes to all recorded Auger spectra, which hinders
our analysis. A larger peak of the hBN spectrum (red curve) as compared
with the bare Ir spectrum (green curve) in the region indicated by
arrow 1 is ascribed to the presence of B in hBN. The excess of B can
also be read out from the hBN spectrum peak in the area indicated
by arrow 2. The Auger spectrum of B\textsubscript{b} (blue curve)
is more similar to bare Ir spectrum, and the two spectra are in principle
indistinguishable in the region of arrow 1. However, a difference
between them is found in the region of arrow 2 where the discernible
peak of the B\textsubscript{b} region is visible. Although small,
this difference between bare Ir and B\textsubscript{b} spectra was
systematically found for different sets of spectra, and we therefore
ascribe it to the presence of B in B\textsubscript{b} regions. One
should keep in mind the concentration of B within the B\textsubscript{b}:
it is approximately $1/12$ of the concentration within the hBN island
(under the assumption of one B atom per B\textsubscript{b} unit cell),
thus diminishing the intensity of the B Auger peaks by the same factor.

The disintegration of hBN and the presence of additional B species
on the surface is complemented by several published studies of hBN/Ir(111).
From their XPS results, Usachov \textit{et al}. found that B 1s and
N 1s spectra intensities decrease rapidly if the sample is kept at
temperatures higher than 1000 \textdegree C \cite{Usachov2012}, ascribing
this to hBN decomposition accompanied by N desorbtion and B diffusion
to the Ir subsurface regions. Moreover, the authors find, besides
the main peak assigned to hBN, two additional B peaks that signify
the existence of two additional forms of boron. In the same study,
LEED images indicate a $\left(6\times2\right)$ superstructure in
addition to the moiré diffraction spots, which is evidence for the
presence of B\textsubscript{b} islands. Farwick zum Hagen \textit{et
al.} provided an atomistic view of B\textsubscript{w} islands via
STM that exhibit a $\left(1\times1\right)$ structure \cite{FarwickzumHagen2016}.
There, the existence of B\textsubscript{w} islands was ascribed to
the decomposition of hBN because the authors observed a similar structure
in the inner parts of hBN islands. Therefore, the process of hBN disintegration
as well as the formation of different B structures on the Ir surface
has been evidenced by several experimental techniques and should be
taken into account for optimal hBN synthesis, and considered as initial
steps towards potential synthesis of borophene on iridium.

\subsection{Island shape and alignment with iridium}

\begin{figure*}
\begin{centering}
\includegraphics{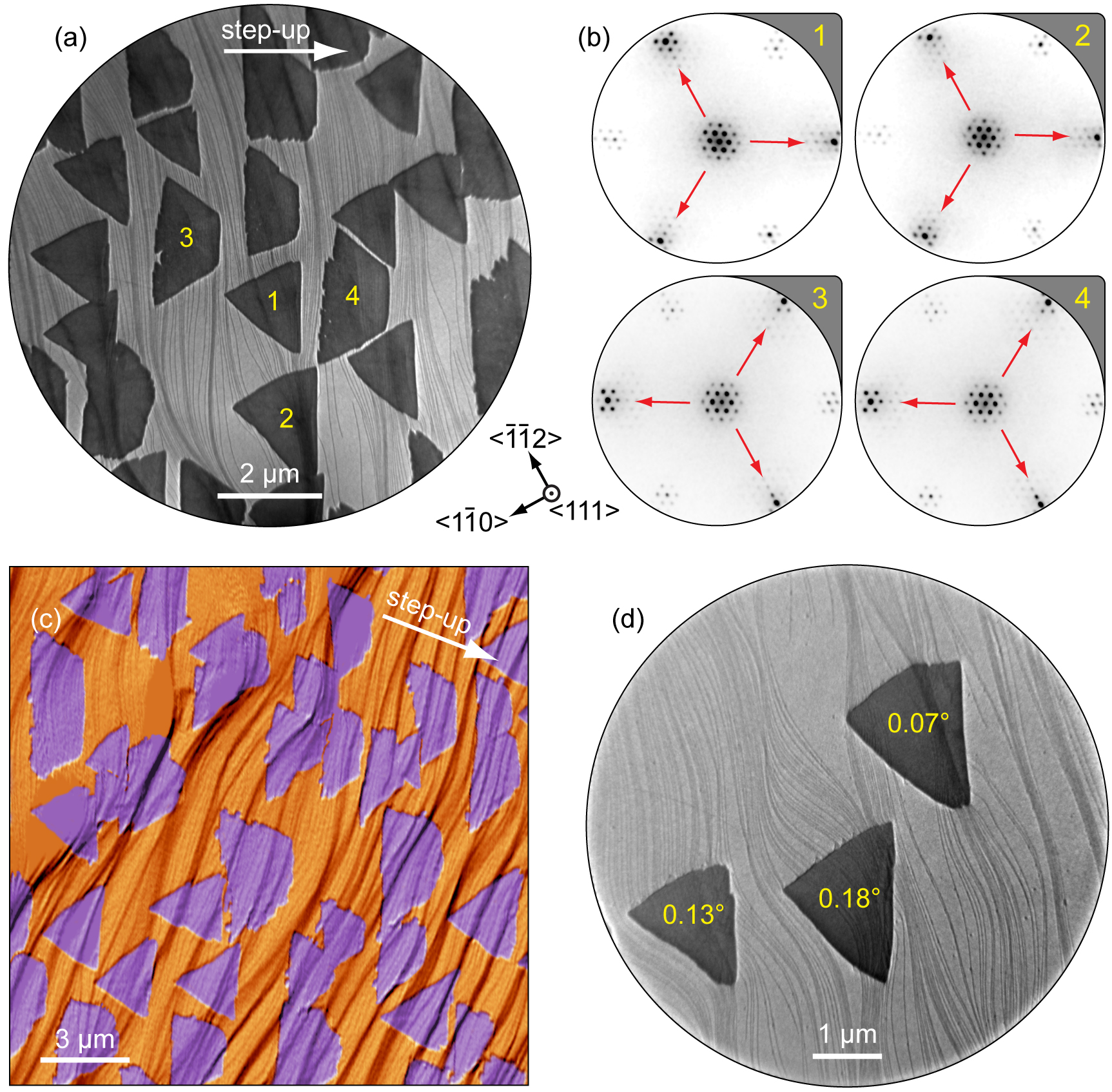}
\par\end{centering}
\caption{\label{fig4}(a) LEEM image of hBN islands on Ir(111). Two island
shapes, triangular and trapezoidal, are distinguished. Thin wavy lines
extending vertically are Ir steps. An arrow points in the step-up
direction of Ir surface. E = 16.2 eV. (b) \textgreek{m}-LEED patterns
of four hBN islands marked in panel (a), enabling identification of
their relative 180\textdegree{} rotation. Arrows point to the groups
of diffraction spots exhibiting higher intensity. E = 40 eV. (c) AFM
image of hBN islands on Ir(111). A first derivative of topography
is shown and hBN islands (along with some B\protect\textsubscript{w}
regions) have been colored in purple for clarity. Wavy lines are Ir
steps; an arrow points in the step-up direction. (d) LEEM image of
three triangular hBN islands exhibiting the noted rotations with respect
to Ir, as inferred from the analysis of their \textgreek{m}-LEED patterns
(not shown). Ir steps are visible as thin wavy lines. E = 17.1 eV.
Crystallographic directions of Ir noted in the center of the figure
apply to all panels. All images were recorded at room temperature.}
\end{figure*}

We now turn to a more detailed analysis of individual hBN islands.
After a CVD synthesis at 900 \textdegree C, two basic island shapes
were found on Ir surface: triangular and trapezoidal, as shown in
Fig. \ref{fig4}(a). This is particularly the case in the initial
stages of growth when the islands are small and their shape (i.e.,
edge straightness and angles between the edges) is not much affected
by the Ir step morphology. Typical island size ranges from 1 to 2
\textgreek{m}m and sides of neighboring triangles and trapezoids are
parallel or nearly parallel, indicating alignment with respect to
the substrate and the existence of a preferred edge type. The island
densities of both triangular and trapezoidal islands were roughly
equal. Moreover, the two island types are rotated by 180\textdegree{}
with respect to each other, as is inferred from their diffraction
patterns shown in Fig. \ref{fig4}(b).

High-symmetry crystal directions, which are easily determined in LEED,
can be mapped to real-space directions in LEEM images, and this allows
for the determination of the hBN island edge type. It is found that
all edges (for both triangles and trapezoids) are of the zig-zag (ZZ)
type under the assumption that no edge reconstruction takes place
(as has been proposed theoretically for hBN on metals \cite{Zhang2016a}).
Indeed, the ZZ edge type has also been confirmed in STM experiments,
and boron-termination has been suggested based on the DFT calculations
\cite{Liu2014d,FarwickzumHagen2016}. Deviations from ZZ edges occur
when islands grow bigger and initial straight edges, mutually connected
at 60\textdegree{} or 120\textdegree , are lost. The mere observation
of triangular island shapes also suggests a ZZ edge type because ZZ
edge configuration in hexagonal 2D materials with bi-elemental basis
enables the presence of one atomic species on the edge, the one which
provides an edge of lower energy. In the case of G growth, initial
small islands are often hexagonal because of mono-elemental basis.

The existence of two 180\textdegree -rotated variants of hBN islands
on Ir(111) was reported before, including the synthesis recipe enabling
prevalence of one orientation over the other \cite{Orlando2014,FarwickzumHagen2016}.
Also, triangular islands of two orientations have been reported for
hBN on strongly interacting substrates such as Ni(111) \cite{Auwarter2003},
Rh(111) \cite{Dong2010}, Ru(0001) \cite{Lu2013} and Co(0001) \cite{Orofeo2013}.
Theoretical calculations for systems without the moiré structure showed
that the two orientations stem from the two possible stable configurations
of hBN on metals \cite{Grad2003a,Laskowski2008}. In both of them,
N atoms are located on top of the Ir atoms (N\textsubscript{top}),
while B atoms can be either in the hcp (B\textsubscript{hcp}) of
fcc (B\textsubscript{fcc}) sites, leading to the formation of two
180\textdegree -rotated domains. However, for systems that develop
the moiré structure, such as hBN/Ir(111), the registry between N and
B atoms with respect to Ir atoms changes across the moiré unit cell.
But even in such cases, two 180\textdegree -rotated orientations of
the hBN layer are preferred, especially if island edge interaction
with the substrate is considered to be dominant in total binding \cite{Camilli2014,FarwickzumHagen2016}.

One of our AFM measurements is shown in Fig. \ref{fig4}(c) where
individual triangular and trapezoidal hBN islands as well as Ir steps
are visible. The short base of the trapezoidal island always forms
in the step-up direction of Ir. It should be noted that because the
trapezoidal islands have two ZZ edges that are parallel, they must
contain different types of edge atoms. Therefore, trapezoidal islands
form a ZZ edge of the opposite composition in the step-up direction
as compared with all other edges. Because this behavior is absent
for triangular islands, we speculate that it is a result of an energetic
balance between the two ZZ edge types (N- and B-terminated) and different
interaction of the two island types with the substrate, in particular
with the substrate step edges. Indeed, STM measurements \cite{FarwickzumHagen2016}
have shown that both island types are triangular in the initial phases
of growth when their dimensions do not exceed typical Ir terrace widths,
indicating the importance of Ir steps for the overall island shape.
The issue of the binding of hBN islands to Ir will be addressed in
more detail in the next section.

The rotational registry of hBN islands synthesized at 900 \textdegree C
was investigated by \textgreek{m}-LEED analysis. With the aid of the
moiré structure, relative rotations of hexagonal 2D overlayers with
respect to Ir are effectively magnified \cite{NDiaye2008} and can
therefore be determined with high precision. For hBN on Ir(111), this
magnification is $\sim11.1$ times, as can be calculated from the
lattice constant values given in Subsection 3.1. By scanning over
many islands, small rotations in the sub-degree region were found.
An example is given in Fig. \ref{fig4}(d) for three neighboring islands
exhibiting small rotations (0\textdegree{} is defined as an alignment
of hBN ZZ direction and dense packed atomic rows of Ir). A small angle
misalignment between hBN and Ir lattices was reported before for hBN/Ir(111)
\cite{Schulz2014,FarwickzumHagen2016} and is also known to be present
in G on Ir(111) \cite{NDiaye2008,Coraux2008}. The major factor influencing
orientational quality of both 2D materials is the synthesis temperature
\cite{Hattab2011,Usachov2012}, and in our case it could be improved
at the expense of the B\textsubscript{b} and B\textsubscript{w}
regions formation.

\subsection{Work function and interaction with iridium}

\begin{figure}
\begin{centering}
\includegraphics{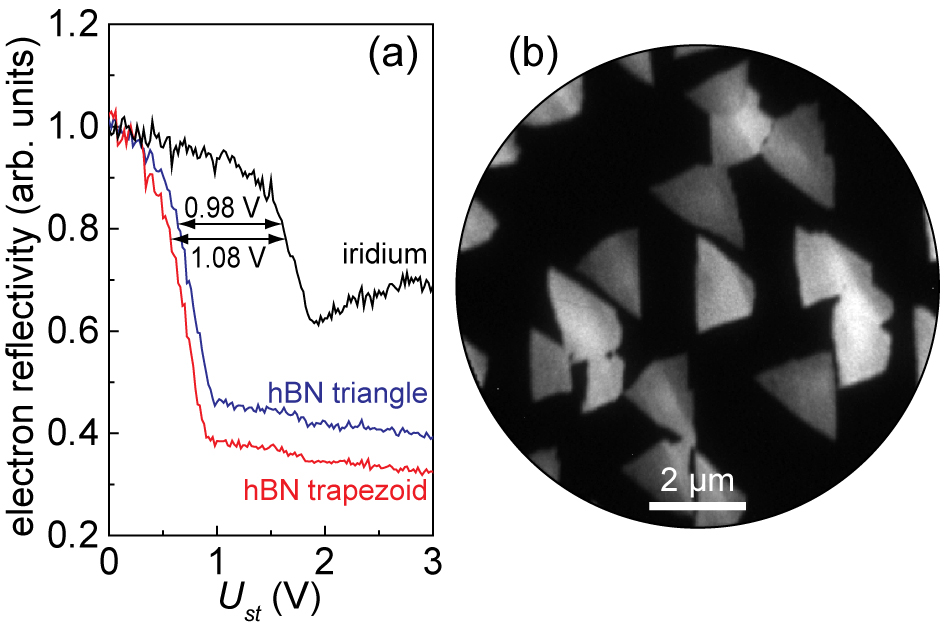}
\par\end{centering}
\caption{\label{fig5}(a) Bright field IV-LEEM spectra taken form Ir (black
curve) and the two types of hBN islands (blue and red curves). Both
island types display a reduction of the WF as visible from the offsets
in start voltage ($U_{st}$), which are indicated by double-headed
arrows. (b) PEEM image of hBN islands on Ir. Due to the WF modifications,
the Ir surface remains dark while the two island types exhibit different
photoemission intensity (see text for details).}
\end{figure}

Basic information about the electronic structure of the triangular
and trapezoidal islands was obtained from the combination of IV-LEEM
spectroscopy and PEEM. In Fig. \ref{fig5}(a), three normalized bright
field IV-LEEM spectra taken from bare Ir and the two hBN island types
are shown. By looking at the drop in electron reflectivity when increasing
the electron energy, it is obvious that the examined areas have different
work functions (WFs) \cite{Babout1980}. We estimate the difference
in the WF between Ir and triangular hBN island to be 0.98 eV, while
it is larger for the trapezoidal island and amounts to 1.08 eV. In
other words, trapezoidal islands exhibit a WF that is 0.1 eV lower
that the WF of triangular islands. These values were obtained by performing
a linear fit to the reflectivity shoulders and then determining their
relative offsets at a reflectivity value of 0.8. By taking the WF
of bare Ir to be 5.76 eV \cite{Strayer1973}, hBN triangles and trapezoids
have WFs of 4.78 and 4.68 eV, respectively.

These values are somewhat lower that the ones reported in an $I\left(z\right)$
spectroscopy study of hBN/Ir(111) by Schultz \textit{et al}. \cite{Schulz2014}.
In the same work, the authors showed that more strongly interacting
areas of the hBN/Ir(111) moiré cell exhibit larger WF. Similarly,
from IV-LEEM measurements of G on Ir(111), Starodub \textit{et al.}
\cite{Starodub2011} found that more strongly bound rotational variant
of G has a higher WF. Therefore, we speculate that of the two hBN
island types, triangular ones are more strongly bound to Ir. A difference
in binding strength, although amounting only to 2.8 meV per hBN unit
cell, has also been suggested based on the DFT calculations \cite{FarwickzumHagen2016}.

The binding inequivalence speculation is further supported by a PEEM
measurement shown in Fig. \ref{fig5}(b). It demonstrates that triangular
and trapezoidal islands exhibit different photoemission intensities,
which in fact reflect differences in the electronic structure of the
hBN/Ir(111) system as a whole. The first factor that can contribute
to this difference is the WF variation. The second factor to be considered
is the coherence of the Ir surface state at the $\Gamma$ point \cite{Niesner2012},
which is the most probable source of photoelectrons in PEEM images,
because it remains present on the surface after hBN formation \cite{Usachov2012}.
A strong hBN-Ir interaction can potentially cause quenching of the
surface state \cite{Schulz2014}, which would reduce the photoemission
yield in PEEM. Therefore, both IV-LEEM and PEEM measurements suggest
stronger interaction of the triangular hBN islands with Ir, as opposed
to the trapezoidal ones. Overall, different binding of the two island
types could be the source of different growth modes and consequently,
in combination with substrate morphology, different island shapes.
Trapezoidal islands have been observed for hBN on Pt(111) where binding
to the substrate is weak, but their origin was not discussed \cite{Zhang2015c}.
In contrast, strongly interacting substrates such as Ru or Rh do not
show signs of trapezoidal islands growth. Therefore, there is an intimate
relation between the strength of hBN-substrate interaction (especially
at the step edges) and the final shape of the islands.

Finally, we note that photoemission intensity variations are visible
within individual islands in Fig. \ref{fig5}(b). Comparison with
the corresponding LEEM images of the same area (not shown) reveals
that this change is related to the step morphology of Ir, more precisely
step bunches. Knowing that the emission angle of the photoelectrons
originating from different areas on the islands does not change sufficiently
to cause such changes, intensity variations are a sign of binding
strength modification within a single island due to the stronger interaction
with Ir step edges, an effect that is known to be present in G on
Ir \cite{Rakic2016}.

\section{Conclusion}

\begin{figure}
\begin{centering}
\includegraphics{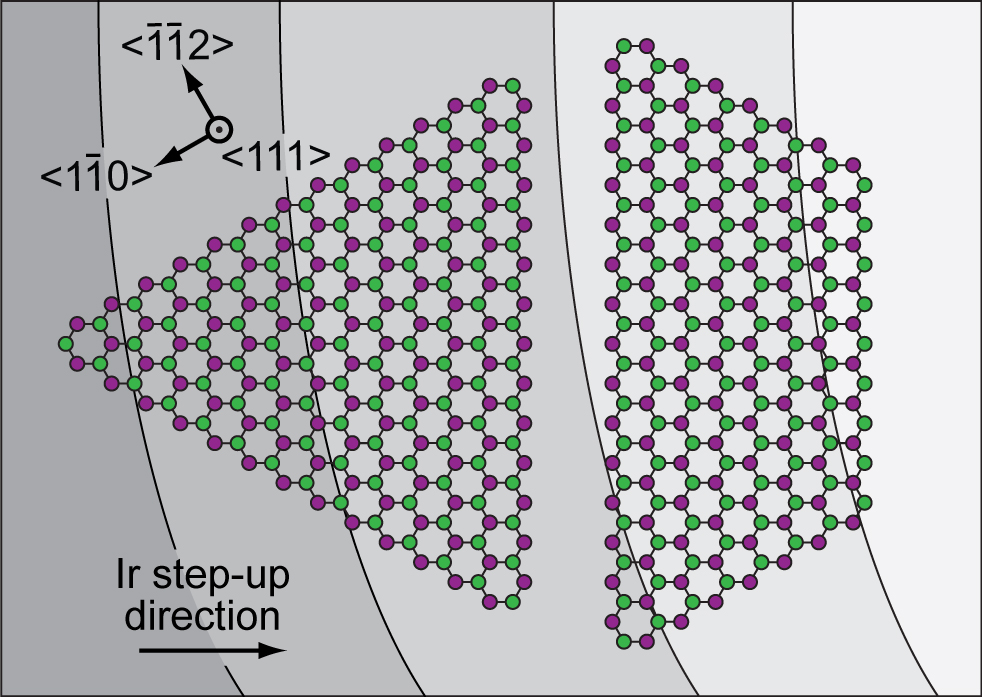}
\par\end{centering}
\caption{\label{fig6}Schematic models of the triangular and trapezoidal hBN
islands found on Ir(111), which are rotated by 180\textdegree{} with
respect to each other. All island edges are of the ZZ type and of
the same composition, except for the short base of the trapezoidal
island. Black lines extending vertically are the Ir step edges. Crystallographic
directions and the step-up direction of Ir are marked in the top-left
and bottom-left corner, respectively.}
\end{figure}

Atomically thin hBN islands of excellent crystal quality and size
up to $\sim10$ \textgreek{m}m can be easily synthesized on the (111)
surface of Ir via CVD process with borazine as a precursor. Two types
of hBN islands have been observed: triangular and trapezoidal, which
are rotated by 180\textdegree{} with respect to each other and which
have ZZ edges. Both island types have the same atom species on all
edges (presumably boron, based on the DFT calculations from Ref. \cite{FarwickzumHagen2016})
except for the short base of the trapezoidal islands where an edge
of opposite composition must exist. These structural results are illustrated
in Fig. \ref{fig6}. Based on photoemission and spectroscopic data,
we conclude that different shapes of the islands originate from different
interactions with Ir. Moreover, Ir steps are crucial for the growth
of hBN because they break the mirror symmetry of the 180\textdegree -rotated
islands, which triggers different growth modes of hBN. This results
in either triangular or trapezoidal islands, and the short base of
the trapezoidal islands always forms in the step-up direction of the
Ir surface. For synthesis temperatures lower than $\approx950$ \textdegree C,
hBN disintegration and dissolution of B in the Ir subsurface areas
are greatly inhibited and hence such temperatures should be used in
order to avoid formation of boron reconstructed regions. In these
regions, boron atoms either fill large areas of Ir terraces (B\textsubscript{b}
regions) or form small irregular patches attached to hBN islands (B\textsubscript{w}
regions). This provides new possibilities for the production of other
novel boron-based 2D materials, such as borophene. Overall, our results
give new insights into CVD synthesis of hBN and can be used to steer
future production of this important 2D material.

\section{Acknowledgments}

Marko Kralj is acknowledged for critical reading of the manuscript
and useful discussions. M.P. would like to thank the Alexander von
Humboldt Foundation for financial support.

\section{References}

\bibliographystyle{apsrev4-1}
\bibliography{references}

\end{document}